\documentclass[conference]{IEEEtran}
\IEEEoverridecommandlockouts
\usepackage{cite}
\usepackage{amsmath,amssymb,amsfonts}
\usepackage{algorithmic}
\usepackage{graphicx}
\usepackage{textcomp}
\usepackage{listings}
\usepackage{xcolor}
\usepackage{booktabs}
\usepackage{verbatim}
\usepackage{url}
\usepackage{float}

\lstdefinestyle{cleanpython}{
  language=Python,
  frame=single,
  captionpos=b,
  breaklines=true,
  breakatwhitespace=true,
  showspaces=false,
  showstringspaces=false,
  showtabs=false,
  tabsize=2,
  stepnumber=1,
  basicstyle=\ttfamily\scriptsize,
  keywordstyle=\color{blue}\bfseries,
  commentstyle=\color{gray}\itshape,
  stringstyle=\color{red},
  backgroundcolor=\color{white},
  morekeywords={self},
}

\lstdefinelanguage{Gherkin}{
  keywords={Feature, Scenario, Given, When, Then, And, But},
  keywordstyle=\color{blue}\bfseries,
  sensitive=false,
  morecomment=[l]{\#},
}

\def\BibTeX{{\rm B\kern-.05em{\sc i\kern-.025em b}\kern-.08em
    T\kern-.1667em\lower.7ex\hbox{E}\kern-.125emX}}
\begin{document}

\title{Exploratory Study on Private GPTs for LLM-Driven Test Generation in Software and Machine Learning Development
\thanks{}
}

\author{\IEEEauthorblockN{\textsuperscript{}Jakub Jagielski}
\IEEEauthorblockA{%\textit{dept. name of organization (of Aff.)} \\
\textit{Ambrosys Gmbh}\\
Potsdam, Germany \\
jakub.jagielski@ambrosys.de}
\and
\IEEEauthorblockN{\textsuperscript{} Consuelo Rojas}
\IEEEauthorblockA{%\textit{Departament de Fisica} \\
\textit{Universitat Politécnica de Catalunya}\\
Terrassa, Spain \\
https://orcid.org/0009-0000-4466-8625}
\and
\IEEEauthorblockN{\textsuperscript{} Markus Abel}
\IEEEauthorblockA{%\textit{dept. name of organization (of Aff.)} \\
\textit{Ambrosys GmbH}\\
Potsdam, Germany \\
markus.abel@ambrosys.de, \\
https://orcid.org/0000-0001-8963-6010}

}

\maketitle

\begin{abstract}
In this contribution, we examine the capability of private GPTs to automatically generate executable test code based on requirements. More specifically, we use acceptance criteria as input, formulated as part of epics, or stories, which are typically used in modern development processes. This gives product owners, or business intelligence, respectively, a way to directly produce testable criteria through the use of LLMs. We explore the quality of the so-produced tests in two ways: i) directly by letting the LLM generate code from requirements, ii) through an intermediate step using Gherkin syntax. As a result, it turns out that the two-step procedure yields better results - where we define better in terms of human readability and best coding practices, i.e. lines of code and use of additional libraries typically used in testing. Concretely, we  evaluate prompt effectiveness across two scenarios—a simple “Hello World” program and a digit classification model—showing that structured prompts lead to higher-quality test outputs.

\end{abstract}


\section{Introduction}

As Large Language Models (LLMs) gain traction in software development, their potential to speed up development by automating tasks is gaining increasing interest \cite{survey}. The quality of code generated from prompts is currently a hot topic. In this study we aim to push LLM technology, with a focus on Retrieval Augmented Generation (RAG), to the limit. We emphasize that all models investigated were private, being aware that commercial LLMs, like Claude or ChatGPT are more performant than private ones.

This study investigates how structured acceptance criteria can be paired with LLMs to generate test sets for software projects. We do this by first translating natural language (NL) to an intermediate language with a specific structure as Gherkin Syntax (GS), and using this language to prompt the LLM. We then show how the quality of generated test code improves.
We compare the performance of different types of prompts across two benchmark scenarios: (1) a classic “hello world” program representing a computer program with minimal functionality, and (2) a digit classification machine learning model, reflecting a more realistic use case.

\section{Implementation Overview}
RAG is a technique that improves the performance of LLMs by combining them with external information retrieval systems\cite{b2}. RAG allows the model to search and retrieve relevant documents or data from outside sources, instead of relying on the knowledge embedded within the model itself.
This approach is especially useful for tasks which require highly specific or up-to-date domain knowledge. RAG does not require retraining the LLM on the data, instead it lets the model fetch and incorporate the information on the fly.

\begin{figure}[htbp]
\centerline{\includegraphics[width = 0.5\textwidth]{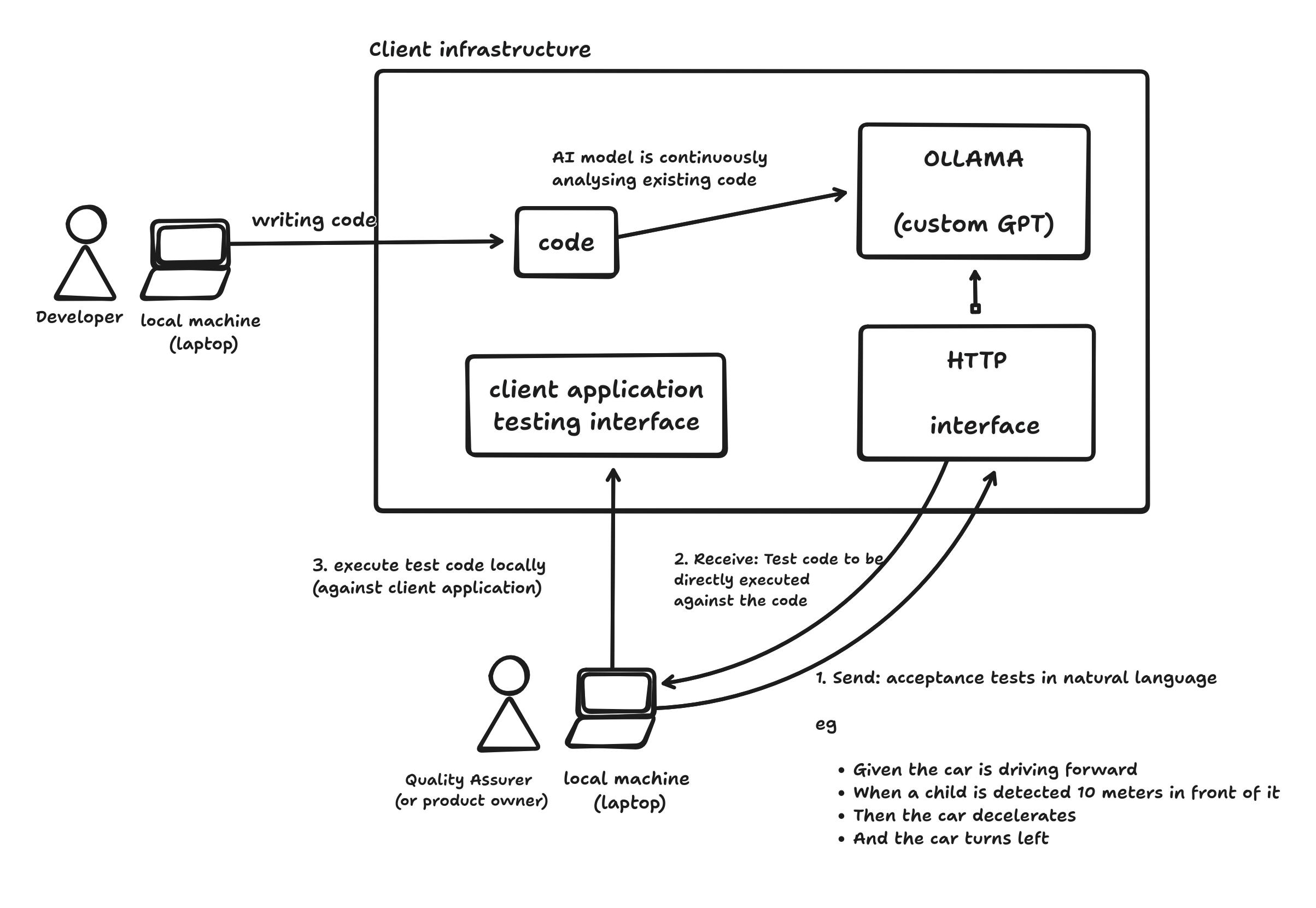}}
\caption{Architectural setup of the system. It shows the client infrastructure and how the different actors interact with the system.}
\label{fig:setup}
\end{figure}

The physical setup of the system is described in Fig. \ref{fig:setup}. The LLM model is deployed on a server and is fronted by an HTTP interface. The workflow is as follows:

\begin{enumerate}
    \item A developer submits code to the server.
    \item A quality assurer prompts the model with acceptance criteria.
    \item The developer’s code is loaded into the model’s context.
    \item The server responds to the quality assurer with the prompt’s result, namely the test code to be executed.
    \item The resulting test code is executed against the developer’s code, to check whether the acceptance criteria have been met. 
\end{enumerate}

\section{Case Studies}
In the following section, we present the two case studies discussed earlier with their results. The codes were given as context to the deepseek-coder-v2 LLM deployed on a server. To evaluate the outcomes, we use the following metrics:

\begin{itemize}
    \item \textbf{Executable (\%):} Percentage of files with one or more executable tests.
    \item \textbf{Pass Rate (\%)} Percentage of the files that have, at least, one executable test that pass.
    \item \textbf{Coverage (\%):} This metric represents the mean percentage of code lines executed during testing, considering only files with at least one executable test. Coverage is calculated as the ratio between the number of lines executed by the tests and the total number of executable lines in the original file.
\end{itemize}

Note: For the sake of readability, we omit inline comments in the code examples provided throughout this paper.

\subsection{Case Study I: Hello World Program}\label{case1}
The first benchmark is a ''Hello World'' program (Listing \ref{lst:hw1}). Upon program execution, the program is expected to log ''hello world'' to the standard output. We started off given the following NL prompt: \textit{''Write pytest tests for the following acceptance criteria: should return hello world''}. The prompt was executed a hundred times the summary of the results can be seen in Table \ref{tab:hw1}.

\begin{figure}[h]
\centering
\begin{minipage}{0.9\columnwidth}
\begin{lstlisting}[style=cleanpython, language=Python, caption={Hello World benchmark program}, label={lst:hw1}, captionpos=b]
def main() -> None:
    print("hello world")

if __name__ == "__main__":
    main()
\end{lstlisting}
\end{minipage}
\end{figure}

Examining the code, only fundamental Python knowledge is sufficient to see that the resulting code cannot possibly work, as in some cases, the function in question is not executed within the test code as is shown in the samples Listing \ref{lst:hw2}.

\begin{figure}[h]
\centering
\begin{minipage}{0.9\columnwidth}
\begin{lstlisting}[style=cleanpython, language=Python, caption={Samples of response to ''Hello World'' Case Study with NL prompts.}, label={lst:hw2}, captionpos=b]
# ============= EXAMPLE 1 ============

import sys
import os
sys.path.insert(0, os.getcwd())  
from main import main  
import pytest

def test_main():
    captured = capfd.readouterr()  
    assert "hello world" in captured.out, f'Expected to see "hello world", but got {captured.out}'
    
# ============ EXAMPLE 2 =============

import pytest
from main import main  

def test_should_return_hello_world():
    captured = capsys.readouterr()  
    assert "hello world" in captured.out  

\end{lstlisting}
\end{minipage}
\end{figure}

Next, we repeat the experiments with a structured approach. We hypothesize that writing the prompt in a more structured language might yield better results. The prompt was modified to be \textit{''Write pytest tests for the following acceptance criteria: When I run the main function, Then I receive 'hello world'  in the standard output''} following the GS. This syntax is a structured, human-readable language used to describe software behaviors and requirements in a way that both technical and non-technical stakeholders can understand \cite{b3}. In these experiments, we added the GS file directly to the context alongside the prompt.

The results showed that GS prompt performs better in comparation to NL approach. While the generated test code was sometimes overly complex, it typically passed, proving that re-structuring the prompt into a specific format can lead to better results. Examples of the test code generated using GS prompts are provided in Listing \ref{lst:hw3}.

\begin{figure}[h]
\centering
\begin{minipage}{0.9\columnwidth}
\begin{lstlisting}[style=cleanpython, language=Python, caption={Samples of response to ''Hello World'' Case Study with structured prompt.}, label={lst:hw3}, captionpos=b]
# ============= EXAMPLE 1 ============

import sys
from io import StringIO
import pytest

@pytest.fixture(autouse=True)
def redirect_stdout():
    old_output = sys.stdout
    new_output = StringIO()
    sys.stdout = new_output
    yield  
    sys.stdout = old_output


import hello_world.main as m
def test_main():
    capturedOutput = StringIO()         
    sys.stdout = capturedOutput        
    sys.stdout = sys.__stdout__       
    assert capturedOutput.getvalue().strip() == "hello world" 

# ============= EXAMPLE 2 ============ 

import pytest

import hello_world.main as m
def test_main(capsys):  
    m.main()             
    captured = capsys.readouterr()  
    assert "hello world" in captured.out    

\end{lstlisting}
\end{minipage}
\end{figure}

Table \ref{tab:hw1} presents the core metrics from the Hello World benchmark, comparing test suites generated using NL and GS structured prompts.

The table shows a significantly higher percentage of GS-generated test scripts are executable compare to those from NL prompts, with a difference over 25\%. This reflects that GS prompts generate more syntactically and structurally complete test cases, though they often show unresolved external dependencies.

It's important to note that the Pass Rate is dramatically different, while only 15\% of the files generated pass, at least, one test, nearly 97\% of GS generated files succeed once the post-processing. Suggesting that the main failure for GS test comes from missing imports or dependencies errors, which are easily fixable. In contrast, NL generated test often fails semantically. For example, the call non-existent functions or rely on undefined behavior, which are more difficult to correct without retraining models, re-prompting or the addition of additional context, such as documentation.

Upon a manual inspection of the failing test reveals two main categories of errors:
\begin{itemize}
    \item \textbf{Dependency errors:} or import errors. Affects, primarly, to GS generated tests. These are typically due to missing context, such as folder structure or required libraries. They can be solved in post processing, adding the missing importations and the necessary paths to the original code. LLMs can manage this error by themselves with this information given in the prompt or in the context.
    \item \textbf{Semantic errors:} common in NL-generated test, include undefined methods calls, incorrect assumptions about functions and misused objects. This errors persist beyond post-processing.
\end{itemize}

 Table \ref{tab:hw1} shows that the mean coverage is similar for both NL and Gherkin prompts. However, it is important to note that the NL results include only 73\% of the generated test files, whereas the Gherkin results cover 97\%.

\begin{table}[!htbp]
\caption{Metrics for Case Study I (Sec.~\ref{case1})}
\label{tab:hw1}
\begin{center}
\begin{tabular}{|c|c|c|}
\hline
\textbf{Metric} & \textbf{NL} & \textbf{Gherkins} \\
\hline
Executable (\%) & 71.0 & 97.81 \\
\hline
Pass Rate (\%) & 15.0 & 96.71 \\
\hline
Coverage (\%) & $63.18 \pm 9.94$ & $62.63 \pm 11.48$\\
\hline
\end{tabular}
\end{center}
\end{table}

\subsection{Case Study II: Digit Recognition}\label{sec:case2}

%\textcolor{red}{read and edit to describe table II}
%\textcolor{blue}{lack of error analysis, types and failure rates, prompt effectiveness (not that important as we are not testing in multiple llms)}

For these case study, we trained a a simple feedforward digit classifier using TensorFlow on the MNIST dataset\cite{b4} (Listing \ref{lst:dc1}). We then prompted the RAG to produce a test with NL instructions: \textit{''Write me a pytest to ensure that the classification accuracy of the model is 95 percent''}. As in the ''Hello World'' experiment, this prompt was evaluated 100 times.

\begin{figure}[h]
\centering
\begin{minipage}{0.9\columnwidth}
\begin{lstlisting}[style=cleanpython, language=Python, caption={Feedforward Neural Network for Digit Recognition}, label={lst:dc1}, captionpos=b]
import ...

model_path = constants.MODEL_DIGIT_RECOGNITION_PATH
model = tf.keras.models.load_model(model_path,
                                    compile=False)

class ClassifyDigits(interfaces.IClassifyDigits):
    def __call__(self, images: NDArray) -> NDArray[np.int_]:

        images = images / 255.0                 
        images = images.reshape(-1, 28 * 28)   

        predictions = model.predict(images)
        return np.array([int(np.argmax(prediction)) for prediction in predictions])

if __name__ == "__main__":
    parser = argparse.ArgumentParser()
    parser.add_argument("image_path", type=str)

    args = parser.parse_args()
    x=PIL.Image.open(args.image_path).convert('L').
        resize((28, 28))
    images = np.array(x)
    print(ClassifyDigits()(images=images))
    
\end{lstlisting}
\end{minipage}
\end{figure}

An example of a test generated by the NL prompt is shown in Listing \ref{lst:dc2}. We can see that is a well documented code and includes the instructions on how to change it, so it can be run but as it is this test is not executable, which beats the propose of this research.

\begin{figure}[h]
\centering
\begin{minipage}{0.9\columnwidth}
\begin{lstlisting}[style=cleanpython, language=Python, caption={Sample of response to ''Digit Recognition'' Case Study with NL prompts}, label={lst:dc2}, captionpos=b]
model = load_model(constants.MODEL_PATH)
classifier = ClassifyDigits()

def test_accuracy():

 train_labels = np.array([0, 1, ..., 9])

    if len(train_labels) == 6000: 
        train_images = np.zeros((len(train_labels), 28, 28)) 

        predicted_labels = classifier(train_images)
        accuracy = sum([1 for i in range(len(predicted_labels)) if predicted_labels[i] == train_labels[i]]) / len(predinted_labels)

        assert accuracy >= 0.95, f"Accuracy is below expected threshold: {accuracy}"
    else:
    pytest.skip("Skipping test as MNIST data size does not match the required sample count.")
    
\end{lstlisting}
\end{minipage}
\end{figure}

For comparison, we prompt the RAG with CS prompt (Listing \ref{lst:dc_gs}). We have to note that this prompt is far more explicit about the coding conventions and indicates that has to use only available resources. Unlike the NL prompt, this describes the clear test scenario, with the expected used resources. This precision leads to better results.

\begin{figure}[h]
\begin{minipage}{0.9\columnwidth}
\begin{lstlisting}[style=cleanpython, language=Gherkin, caption={Gherkin Syntax prompt for Digit Recognition unit test generation}, label={lst:dc_gs}, captionpos=b]
Given the files, generate me a pytest for the acceptance criteria:

Given I have a trained digit recognition model
And  a test set
When I classify the test set using that model
Then  I get an accuracy of more than 95 percent

Use good coding python conventions, and make sure to import any packages / fixture that you use. Do not refer to any non-existent resources. When importing files, take the project structure into account.
\end{lstlisting}
\end{minipage}
\end{figure}

In Table \ref{tab:dc1}, summarize the outcomes of both prompting techniques and it shows that the executability for GS tests is significantly higher (over 90\%) than the NL generated tests (49\%). This highlights the importance of clarity and structure in the prompts. As seen in the ''Hello World'' case study, most of the errors for the NL test units are due to semantic failures (using undefined classes, resources that were not available. These failures are deep and non-trivial to fix. In contrast, GS unit test often failed due to import errors (missing TensorFlow imports, one of the most common), which were easily corrected through post processing.

\begin{table}[h]
\caption{Metrics for Case Study II (Sec.~\ref{sec:case2})}
\label{tab:dc1}
\begin{center}
\begin{tabular}{|c|c|c|}
\hline
\textbf{Metric} & \textbf{NL} & \textbf{Gherkins} \\
\hline
Executable (\%) & 49.43 & 94.56 \\
\hline
Pass Rate (\%) & 9.09 & 6.81 \\
\hline
Coverage (\%) & $50.00 \pm 0.00$ & $50.00 \pm 0.00$ \\
\hline
\end{tabular}
\end{center}
\end{table}

Prompting with GS format also affect how test are constructed. Nearly all test files generated from GS prompts contains fewer cases, this is because GS guide indicates that more than 5 examples affects generalization. Meanwhile, NL generated test, most common failure is hallucination (already mentioned, the use of nonexistent resources). These test tend to completely fail, resulting in a drop in executability. Providing more contextual information would likely reduce these failures. But what is interesting, is that GS prompts were less likely to hallucinations, suggesting that structured format, when pair with contextual grounding reduces the amount of assumptions made by the LLMs.

The mean coverage (as seen in Table \ref{tab:dc1}) for NL and Gherkin is the same. However for NL, only 46\% of test files contained runnable code, whereas for Gherkin it was 94\%.

As it's shown in the Table \ref{tab:dc1} with the results, both prompt styles have very low test pass rates (NL: 9.06\% and GS: 6.81\%), there are several explanations for this:
\begin{enumerate}
    \item Neural networks are inherently stochastic, and the outputs can vary, slightly, across runs. Thus asserting a fixed threshold is not ideal for testing machine learning models.
    \item Some test generated datasets or import the training dataset for the testing. This can lead to data leakage, artificially inflating or deflating performances, resulting in tests that pass or fails inconsistently.
\end{enumerate}

These observations emphasize the limitations of prompt-based test generation for machine learning workflows given the nature of this elements. And highlights the importance of framing test prompts in a more conductual way, making sure instances are working and not the results of these models.

%\textcolor{red}{WE HAVE TO ADDRESS THAT THIS ARE MOSTLY TESTING IN THE TRAINING SET, SO THE TEST IS BIAS AND WE HAVE TO SPECIFY THAT WE NEED 95\% IN THE TEST DATASET}

\section{Gherkin Syntax Prompt Generator}

We discovered experimentally that prompting the LLM using a GS yields higher-quality test code. Following from that, we investigated whether RAG technology can be used to streamline the test generation workflow.
The idea is to let the LLM handle the whole process. The quality assurer submits acceptance criteria in natural language, the RAG converts the input to GS, and the LLM is re-prompted using the RAG generated prompt.
Gherkin documentation at https://cucumber.io/docs/gherkin/reference/ was exported as a pdf and provided as context to the RAG. The prompt submitted to the LLM was que quote below, and the results are shown in Listing \ref{lst:ghk_g1}.\\

\begin{quote}
\textit{
Given the Gherkin reference, generate me Gherkin syntax-compliant instructions for the following acceptance criteria:
\begin{itemize}
    \item the function should log hello world to the console.
    \item the function should recognize more than 95 percent of digits correctly.
\end{itemize}
Write them in an explicit way to be easily testable
}
\end{quote}

\begin{figure}[h]
\centering
\begin{minipage}{0.9\columnwidth}
\begin{lstlisting}[style=cleanpython, language=Gherkin, caption={Generated Gherkin Syntax from natural language input}, label={lst:ghk_g1}, captionpos=b]
Feature: Functionality Requirements

  Scenario: Logging "hello world" to the console
    Given a function that logs messages to the console
    When I call this function with no arguments or parameters
    Then it should log "hello world" to the console

  Scenario: Recognizing more than 95 percent of digits correctly
    Given a function capable of recognizing digit strings
    And a dataset containing various single-digit numbers (0 through 9) and other random string data
    When I test this function with at least ten different inputs from my dataset
    Then it should recognize over 95% of the input as digits correctly
\end{lstlisting}
\end{minipage}
\end{figure}

The LLM paired with the RAG can successfully generate Gherkin syntax conformant acceptance criteria from natural language. However, in some cases it made up information. In the digits example, the LLM added “various single-digit numbers (0 through 9) and other random string data” which is something we have not instructed the LLM to do.
Next, we attempt to generate tests for benchmark 1 again (the classic hello world program), using the RAG generated prompt. We prompt using the following:\\ 

\begin{quote}
\textit{Using the submitted files, write pytest tests for the following acceptance criteria:}

\textit{Scenario: Logging "hello world" to the console}

\textit{Given a function that logs messages to the console}

\textit{When I call this function with no arguments or parameters}

\textit{Then it should log "hello world" to the console}

\textit{Note: Don’t over-engineer the code; use what is available, such as packages and fixtures.}\\
\end{quote}

The resulting generated test code is given in Listing \ref{lst:ghk_g2}. The observation from this is that using Gherkin generated from acceptance criteria (with RAG) yields better final results than no Gherkin at all, although writing Gherkin by hand still yields the best results.

\begin{figure}[h]
\centering
\begin{minipage}{0.9\columnwidth}
\begin{lstlisting}[style=cleanpython, language=Python, caption={The test code generated by the LLM using its own Gherkin prompt. The prompt was generated by the LLM from a natural language prompt.
}, label={lst:ghk_g2}, captionpos=b]
import sys
from io import StringIO

# Assuming main() function is defined somewhere accessible by this script
def test_logging():
    # Create a buffer to capture print output
    captured_output = StringIO()
    sys.stdout = captured_output # Redirect stdout to the buffer

    from hello_world.main import main # Replace 'your_script' with the name of your script file
    main() # Call the function you want to test

    sys.stdout = sys.__stdout__ # Reset redirect
    assert captured_output.getvalue().strip() == "hello world"
\end{lstlisting}
\end{minipage}
\end{figure}

\section{Conclusion}

This study demonstrate that is feasible generate code test using private GPT models. Our experiments show that prompting in a structured manner generally improve the quality of the tests generated.

A key finding is that interacting with LLMs through a structured language, specially Gherkin syntax, significantly enhances the quality and reliability of the produced test code. Even more, the structured syntax can be automatically generated by the LLM when provided with the specifications as context.

Further exploration would include evaluate the use of a specialized LLM to translate natural language prompts into a intermediate representations like Gherkin, and employ a dedicated LLM, to generate the corresponding code from the structured prompt.

Overall, these approaches guide us toward a more robust and automated test generation workflow. We plan to further investigate the comparative performance of commercial versus non-commercial LLMs under established coding standards and more complex scenarios.

\section*{Acknowledgment}

Part of this work was funded by the European Union, under HORIZON-MSCA-2022-DN, \textit{Improving BiomEdical diagnosis through LIGHT-based technologies and machine learning} ``BE-LIGHT'' (GA nº 101119924 - BE-LIGHT) project.

Views and opinions expressed are however those of the author(s) only and do not necessarily reflect those of the European Union nor UPC. Neither the European Union nor the granting authority can be held responsible for them.

The code and data used in this paper are publicly available at \url{https://github.com/consuelorojas/Exploratory-LLM}.

\end{document}